\documentclass[reprint,amsmath,amssymb,aps,prl]{revtex4-1}
\usepackage{graphicx}
\usepackage{bm}
\begin{document}
\title{Multifractal sets of coherent and incoherent vortices in turbulence}
\author{Susumu Goto}
\email{s.goto.es@osaka-u.ac.jp}
\author{Daiki Watanabe}
\affiliation{
Graduate School of Engineering Science, 
the University of Osaka, Toyonaka, 560-8531, Japan}
\author{Tsuyoshi Yoneda}
\affiliation{Graduate School of Economics, 
Hitotsubashi University, Kunitachi, 186-8601, Japan}
\date{\today}
\begin{abstract}
We numerically verify multifractal theory (Frisch and Parisi 1985) for turbulence using simulation data at a high Reynolds number. First, we propose a simple method to directly estimate the multifractal dimension $D(h)$ of vortical structures with a given H\"older exponent $h$. Thus measured $D(h)$ is in good agreement with indirectly measured experimental data. Then, we demonstrate that these structures for $h\ll1/3$ form the hierarchy of coherent eddies, while those for $h\gg1/3$ are featureless.
\end{abstract}
\maketitle

{\it Introduction.} Fractal \cite{Mandelbrot-1982} is an important concept that appears in various systems. Since fractal sets are self-similar without a characteristic length, the correlation function of a quantity characterizing them obeys a power law whose exponent is related to the fractal dimension \cite{Falconer-2014}. Turbulence is often introduced as a representative fractal system because we imagine that vortices of various sizes form a self-similar hierarchy on the inertial length scales, which are defined as those sufficiently shorter than the velocity correlation length $L$ characterizing external forces or boundary conditions, yet sufficiently longer than the viscous length $\eta$. 

In order to characterize the hierarchy of vortices, we quantify the swirling velocity of a vortex of size $r$ using the difference, $\Delta u(r)$, in longitudinal velocity between two points separated by distance $r$, and define the $p$-th order longitudinal velocity structure function,
\begin{equation}
\label{eq:S_p-def}
 S_p(r)=\overline{[\Delta u(r)]^p}
 \quad(p=1,2,\ldots)
 \:,
\end{equation}
as the representative statistics of vortices on scale $r$. Here, $\overline{\ \cdot\ }$ denotes the spatial average. If turbulence is fractal, $S_p(r)$ obeys a power law. Indeed, this has been verified by many experiments and numerical simulations (see figure 12 of Ref.~\cite{Dubrulle-2019}, for example). In particular, the $2/3$ power law for $p=2$,
\begin{equation}
\label{eq:S_2}
 S_2(r)\sim\overline{\epsilon}^\frac23r^\frac23
 \quad(\eta\ll r\ll L)
 \:,
\end{equation}
predicted by Kolmogorov \cite{Kolmogorov-1941} using the mean dissipation rate $\overline{\epsilon}$ of kinetic energy per unit mass, is equivalent to the well-known $-5/3$ power law,
\begin{equation}
\label{eq:E-5/3}
 E(k)
 =
 K\overline{\epsilon}^{\frac23}k^{-\frac53}
 \:,
\end{equation}
of the energy spectrum in the wave-number range, $2\pi/L\ll k\ll2\pi/\eta$. In (\ref{eq:E-5/3}), $K$ is a universal coefficient called the Kolmogorov constant. A numerous number of studies support (\ref{eq:E-5/3}) and the universality of $K$ \cite{Sreenivasan-1995}.

In addition, for the case where $p=3$ in (\ref{eq:S_p-def}), we can derive $S_3=-4\overline{\epsilon} r/5$ from the Navier--Stokes equation \cite{Kolmogorov-1941c}. The experimental support for (\ref{eq:E-5/3}) and the rigorous expression for $S_3$ provide evidence that turbulence is fractal on the inertial scales ($\eta\ll r\ll L$). However, it is difficult to observe concrete self-similar fractal sets of vortices in turbulence.

The difficulty of turbulence research stems from its strong nonlinearity, which leads to energy exchange between scales. Richardson wrote a verse \cite{Richardson-1922} to briefly describe the energy transfer from large to small scales in turbulence. This depiction, known as the energy cascade, led, through Kolmogorov's \cite{Kolmogorov-1941} keen insight, to the proposal of his second similarity hypothesis: namely, the statistics in the inertial range are described solely by $\overline{\epsilon}$ regardless of the type of turbulence. Dimensional analysis under this simple hypothesis yields (\ref{eq:E-5/3}) and 
\begin{equation}
\label{eq:S_p-K41}
 S_p\sim(\overline{\epsilon} r)^{\frac p3}
 \qquad
 (\eta\ll r\ll L)
\end{equation}
for the $p$-th order structure function. Furthermore, the energy cascade picture explains the universality of small-scale turbulence, which gives the basis of turbulence models \cite{Kim-2024} for the computational fluid dynamics across various fields, because the information of large-scale structures is likely lost through the cascade. 

However, Kolmogorov's similarity hypothesis does not hold strictly. Landau \cite{Landau-1959} pointed out that a description based only on the mean dissipation rate $\overline{\epsilon}$ was incorrect and temporal fluctuations must be taken into account. In addition, Kolmogorov himself \cite{Kolmogorov-1962} proposed a modified theory taking into account spatial fluctuations, which we call the spatial intermittency, in the energy dissipation rate. Since then, the issue of spatial intermittency \cite{Kraichnan-1974,Frisch-1995,Buaria-2026} has been actively studied, particularly to evaluate deviations from (\ref{eq:S_p-K41}).

It is ironic that the energy cascade picture underlying Kolmogorov's similarity hypothesis \cite{Kolmogorov-1941} explains why energy transfer at smaller scales, and therefore the dissipation rate, become spatially intermittent, resulting in the violation of the hypothesis (compare Figs.~7.2 and 8.9 in Ref.~\cite{Frisch-1995}). Although (\ref{eq:S_p-K41}) provides a good approximation for small $p$, the turbulence statistics for large $p$ cannot be accurately described without considering the spatial intermittency of the energy dissipation rate. The so-called cascade models such as the $\beta$ model \cite{Frisch-1978} were proposed to describe the correction to the Kolmogorov theory by quantifying this intermittency due to the energy cascade. 

In this study, we introduce a new perspective on the intermittency in turbulence. In contrast to previous studies, which focused primarily on statistical quantities such as the power-law exponent $\zeta_p$ in the inertial range of $S_p$ defined as (\ref{eq:S_p-def}), the objective of the present study is not to quantify correction to the Kolmogorov law, but to elucidate the underlying physics. We investigate the spatial structures of vortices to establish a concrete image of their fractal feature.

{\it Multifractal description of turbulence.} Let us proceed to the formulation of the problem. First, we recall that Kolmogorov's similarity hypothesis gives a good approximation at least for lower-order quantities. As mentioned above, according to his second similarity hypothesis in the inertial range, the turbulence statistics are determined by $\overline{\epsilon}$, and (\ref{eq:S_p-K41}) holds. Therefore, for $p=1$, the hypothesis leads to
\begin{equation}
\label{eq:h=1/3}
 \overline{\Delta u}\sim\overline{\epsilon}^\frac13 r^\frac13
 \qquad
 (\eta\ll r\ll L)
 \:.
\end{equation}
%

Since (\ref{eq:h=1/3}) represents the fractal nature of turbulence, the idea by Mandelbrot \cite{Mandelbrot-1974} that spatial intermittency can be captured through a generalization of fractal seems appropriate. The multifractal analysis is the framework for this purpose, which has been applied not only to turbulence but also to various fields such as medical science \cite{Ivanov-1999}, finance \cite{Mandelbrot-1997}, geophysics \cite{Schertzer-1987, Hirabayashi-1992}, network theory \cite{Feldmann-1998}, and linguistics \cite{Rosmawati-2024}. It was Frisch and Parisi \cite{Parisi-1985} who first formalized this picture
for turbulence, where (\ref{eq:h=1/3}) is generalized with the H\"older exponents $h$ as
\begin{equation}
\label{eq:h}
 \overline{\Delta u}\sim r^h
 \:.
\end{equation}
Under Kolmogorov's similarity hypothesis, $h$ is uniformly $1/3$, whereas multifractal theory allows for a distribution of $h$ and defines the fractal dimension $D(h)$ of the subset with a given $h$. The $p$-th order longitudinal velocity structure function (\ref{eq:S_p-def}) is then expressed as
\begin{equation}
 S_p=\int r^{3-D(h)+ph}\:d\mu(h)
 \:,
\end{equation}
where $d\mu(h)$ is the measure representing the distribution of subsets with H\"older exponent $h$. By using the method of steepest descent, the exponent $\zeta_p$ of the power law in the inertial range of $S_p$ is expressed as  
\begin{equation}
\label{eq:zeta_p}
 \zeta_p
 =
 \min_{h}\left(3-D(h)+ph\right)
 \:.
\end{equation}
Since Kolmogorov's similarity implies that $h=1/3$ uniformly in space, i.e.,~$D(1/3)=3$, it follows from (\ref{eq:zeta_p}) that $\zeta_p=p/3$, implying that the multifractal theory includes the monofractal picture by Kolmogorov.

Measurements of $\zeta_p$ were extensively conducted, and there is general agreement that, except for $p=3$ (where $\zeta_3=1$ is guaranteed by the $4/5$ law \cite{Kolmogorov-1941c}), $\zeta_p\ne p/3$. This is referred to as the anomalous scaling in turbulence \cite{Anselmet-1984, Frisch-1995,Buaria-2026}. The cascade models, including the log-normal theory \cite{Kolmogorov-1962}, have focused primarily on predicting $\zeta_p$ that agrees with experimental data. These predictions can be interpreted in terms of $D(h)$. In other words, since $\zeta_p$ and $D(h)$ are linked by the Legendre transform (\ref{eq:zeta_p}), we can determine $D(h)$ from $\zeta_p$. Therefore, each of $\zeta_p$ measured by experiments or predicted by cascade models has corresponding $D(h)$ \cite{Frisch-1995, Meneveau-1987, Meneveau-1991, Sreenivasan-1991}.

{\it Objective.} Despite extensive investigations, the validity of the basic assumption (i.e.,~the set with a given $h$ has fractal dimension $D(h)$) of the multifractal theory is still an open issue \cite{Dubrulle-2019,Sreenivasan-2025}. Several attempts \cite{Chhabra-1989,Nguyen-2019, Dubrulle-2019} to directly measure $D(h)$ are not necessarily conclusive, while studies to verify the assumption through $\zeta_p$ are insufficient because they cannot relate $D(h)$ with spatial structures. Therefore, the main research question of the present study is as follows: {\it is there any concrete substance to multifractal sets in turbulence?} The conclusion is that the sets for $h\ll1/3$ form the hierarchy of coherent vortices, whereas those for $h\gg1/3$ are featureless.

{\it Methods.} The data examined is developed turbulence driven by an artificial force in a periodic cube with side $L_\text{box}$, referred to as 8192-1 in Ref.~\cite{Ishihara-2016}. The turbulence was simulated by a spectral method with $8192^3$ Fourier modes. The Reynolds number based on the Taylor length is approximately $R_\lambda \approx 1700$, and $L/\eta \approx 4300$, indicating the presence of the inertial range ($\eta\ll r\ll L$) of sufficient width. Details of the numerical method can be found in Ref.~\cite{Ishihara-2016}.

We extract multifractal sets from this turbulent flow and evaluate their dimensions. The procedures are straightforward and consist of the following three steps: (i) First, we perform scale decomposition of the turbulent flow field, (ii) then, we extract subsets with H\"older exponent $h$, (iii) and evaluate the capacity dimension of those sets.

For (i), we use a simple method with the Fourier band-pass filter. Let the representative wave number of the $n$-th band, denoted by B$_n$, be $k_n=k_0b^n$, and let the bandwidth, $a$, on the logarithmic scale be a constant. Then, the $n$-th band B$_n$ is defined as the wave-number range such that 
$
 k
 \in
 \left[{k_n}/{\sqrt{a}},\ \sqrt{a}k_n\right).
$

(ii) For the $n$-th band B$_n$, we define a subset for a given value $h$ of the H\"older exponent as follows. First, we recall that the scaling of the enstrophy ${\cal E}(r)$, i.e.,~the squared vorticity, at scale $r$ corresponding to $h$ is expressed as
\begin{equation}
 {\cal E}(r)
 \sim 
 (r^h/r)^2=r^{2h-2}\sim k^{2-2h} 
\end{equation}
because of the scaling (\ref{eq:h}) of the velocity difference. For the $n$-th band B$_n$, we define the volume $V_n$ of the domain where the scale-decomposed enstrophy ${\cal E}_n$ satisfies 
\begin{equation}
\label{eq:Qn-cond}
 \left(\frac{k_n}{\sqrt{a}k_0}\right)^{2-2h}
 < 
 \frac{{\cal E}_n}{\overline{{\cal E}_0}}
 < 
 \left(\frac{\sqrt{a}k_n}{k_0}\right)^{2-2h}
\end{equation}
and the second invariant $Q_n$ of the velocity gradient tensor in B$_n$ takes positive values. In (\ref{eq:Qn-cond}), $\overline{{\cal E}_0}$ is the spatial average of the scale-decomposed enstrophy in the $0$th band B$_0$ (i.e., the hierarchy level on the largest scale).

(iii) The multifractal theory suggests that $V_n(h)$ obeys a power law of $k_n$ as
\begin{equation}
\label{eq:Vn-power}
 V_n(h)\sim{k_n}^{D(h)-3} 
\end{equation}
with an exponent varying with $h$. Thus, we can estimate $D(h)$ from (\ref{eq:Vn-power}).

The parameters that must be determined through the above procedures are $a$, $b$, and $k_0$. Since $k_0$ can be chosen sufficiently small, we set $k_0=2$. Note that $b$ ($>1$) is a parameter that determines the interval between layers. Since it only determines the resolution of the analysis and does not affect the results, we set $b=1.2$. The remaining parameter is $a$, but there is no objective way to determine it. Therefore, here we show results with $a=2$, while the supplementary material presents results with $a=2^{j/2}$ $(j=1,2,3,4)$.

\begin{figure}
\centering
\includegraphics[bb=0 0 648 536, width=0.48\textwidth]{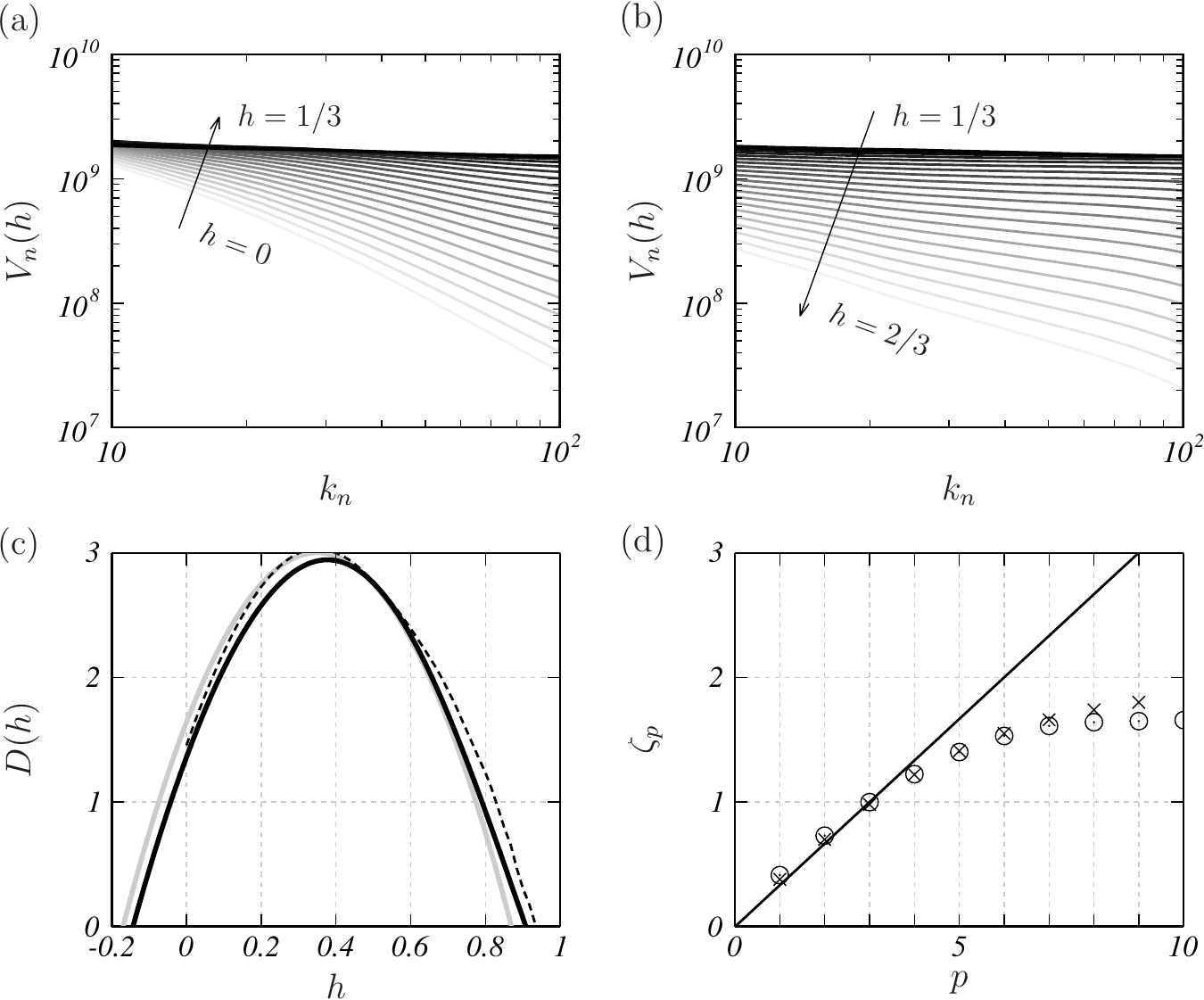}
\caption{(a, b) $V_n$ as functions of $k_n$ for $h=i/60$ (a) $i=0,1,\ldots,20$ and (b) $20,21,\dots,40$. (c) Solid black line, $D(h)$ estimated from the power laws of $V_n$ in (a, b); solid gray line, the parabolic fit suggested in Ref.~\cite{Dubrulle-2019}; dashed line, estimation from the total length of vortex axes. (d) $\circ$, $\zeta_p$ obtained by the Legendre transform of $D(h)$; $\times$, experimental results taken from Figure 13 of Ref.~\cite{Dubrulle-2019}.}
\label{fig:VnDhZp}
\end{figure}

{\it Results and Discussion.} We show the evaluated $V_n(h)$ in Fig.~\ref{fig:VnDhZp}(a, b). According to Ref.~\cite{Ishihara-2016}, the wave-number space of the turbulence examined here is divided into three ranges, and a power law of $E(k)$ deviating from (\ref{eq:E-5/3}) is observed in $5\times10^{-3}\lesssim  k\eta\lesssim2\times10^{-2}$ (i.e.,~$19\lesssim k\lesssim 77$). Focusing on this wave-number range, we can observe power laws of $V_n(h)$ [Fig.~\ref{fig:VnDhZp}(a, b)], and thus $D(h)$ can be evaluated [Fig.~\ref{fig:VnDhZp}(c)]. The result is consistent with previous data of $D(h)$, and $\zeta_p$ evaluated from $D(h)$ through (\ref{eq:zeta_p}) also coincides with experimental results [Fig.~\ref{fig:VnDhZp}(d)]. Although several attempts \cite{Chhabra-1989,Nguyen-2019, Dubrulle-2019} have been made to directly measure $D(h)$ in turbulence, we emphasize again that the method proposed here is extremely simple. 

Although Fig.~\ref{fig:VnDhZp} directly demonstrates that the intermittency of turbulence can be described by the multifractal theory, this is not the main result. In the following, we identify the actual multifractal structures. The method for this is also straightforward. First, we choose $h$, and then we identify the regions satisfying the conditions (\ref{eq:Qn-cond}) and $Q_n>0$ for each B$_n$.

\begin{figure}
\includegraphics[bb=0 0 1800 1200,width=0.48\textwidth]{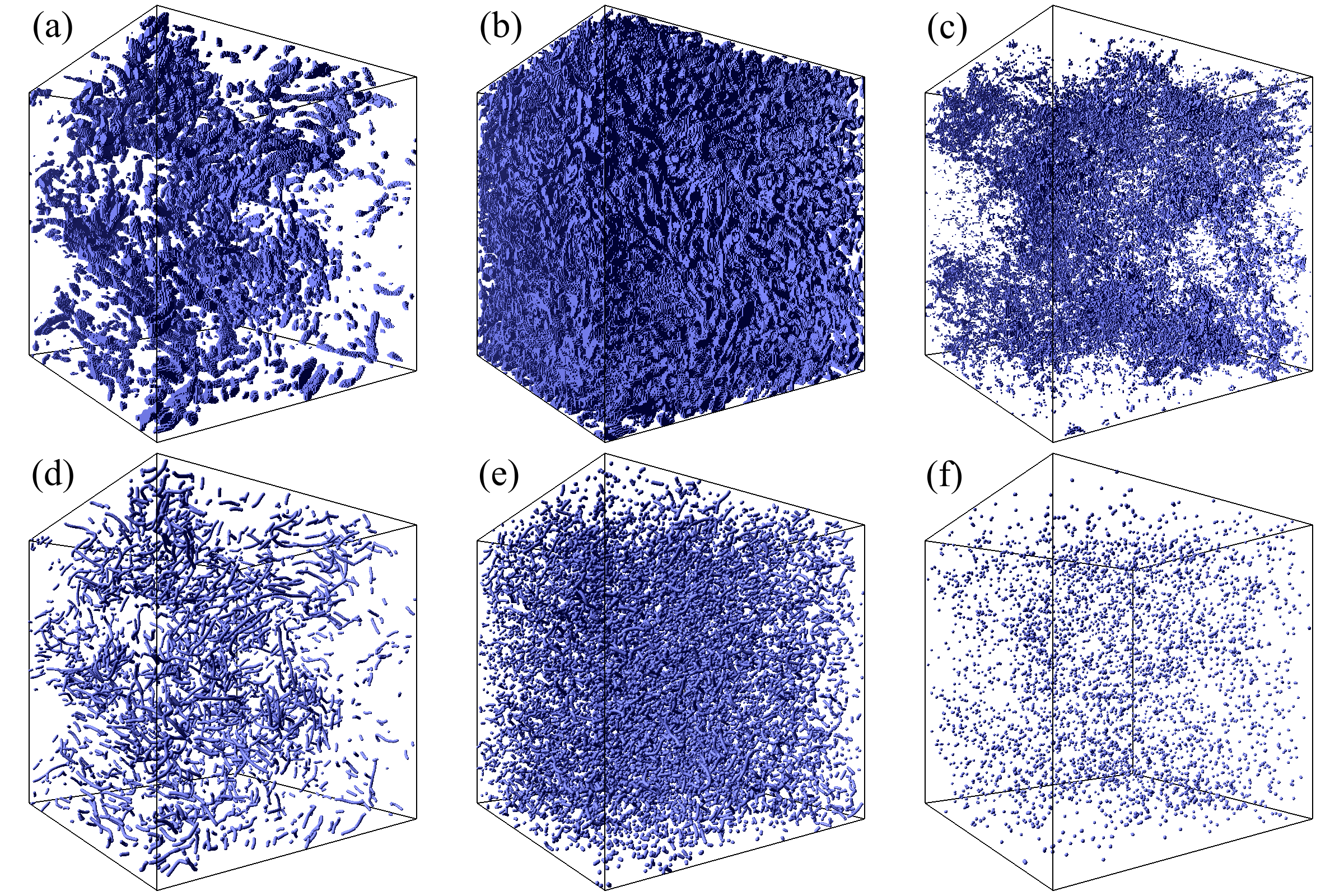}
\caption{(a--c) Visualization of boxes satisfying the conditions (\ref{eq:Qn-cond}) and $Q_n>0$ in the band of $k_n=32$ with H\"older exponents (a) $h=0$, (b) $1/3$, and (c) $2/3$. The size of the box is $(L_\text{box}/2)^3$. (e--f) Vortex axes identified by the low-pressure method for each $h$ corresponding to (a--c), respectively.}
\label{fig:Viz-Q}
\end{figure}

We show in Fig.~\ref{fig:Viz-Q}(a--c) results for $k_n=32$ ($k_n\eta=8.2\times10^{-3}$). It is interesting to observe in Fig.~\ref{fig:Viz-Q} that (a) the structures for $h=0$ form coherent vortex tubes localized in space, (b) those for $h=1/3$ are space filling, and (c) the structures for $h=2/3$ are localized in space but they seem featureless. 

In order to quantify the observed $h$-dependence of the coherence of structures [Fig.~\ref{fig:Viz-Q}(a--c)], we objectively identify vortical structures. Specifically, we extract the axes of vortices at each scale by applying the low-pressure method \cite{Miura-1997} to the scale-decomposed field obtained by the procedure (i) \cite{Goto-2017}. As shown in Ref.~\cite{Tsuruhashi-2022} for turbulence at $R_\lambda \approx 500$, the vortex axes identified in this manner are space filling, and the total length $L_n$ of the vortex axes in band B$_n$ is proportional to $k_n^2$. Since the number of spheres of diameter $r_n$ $(=\pi/k_n)$ required to cover these vortex axes is proportional to $L_n/r_n \sim r_n^{-3}$, their capacity dimension is $3$. Therefore, to capture the multifractal nature, we must impose condition (\ref{eq:Qn-cond}) on the vortex axes to identify subsets of them. Figures \ref{fig:Viz-Q}(d--f) show thus identified subsets for (d) $h=0$, (e) $1/3$, and (f) $2/3$, which correspond to Fig.~\ref{fig:Viz-Q}(a--c), respectively. We can observe a clear qualitative dependence on $h$; namely, coherent tube-like vortices are observed for $h\ll1/3$ [Fig.~\ref{fig:Viz-Q}(d)], but no coherence is observed for $h\gg1/3$ [Fig.~\ref{fig:Viz-Q}(f)]. When $h=1/3$, vortices are space-filling and we can observe moderate coherence [Fig.~\ref{fig:Viz-Q}(e)]. 

\begin{figure}
\centering
\includegraphics[bb=0 0 624 576,width=0.45\textwidth]{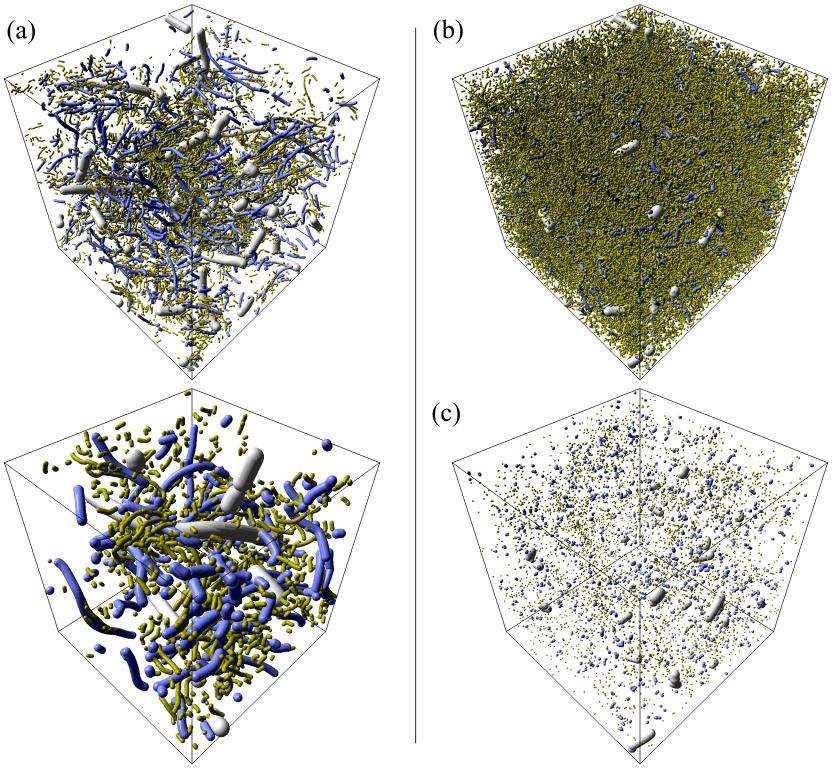}
\caption{Vortex axes in the band of $k_n=4$ (white), $16$ (blue), and $64$ (yellow) with the H\"older exponent (a) $h=0$, (b) $1/3$, and (c) $2/3$ in a cubic domain of size $(L_\text{box}/2)^3$. The bottom plot in (a) is the magnification in a cube of size $(L_\text{box}/4)^3$. Results for other locations are given in the supplemental material.}
\label{fig:axis}
\end{figure}

This $h$-dependence of the coherence and incoherence of vortices is observed regardless of the scale as demonstrated in Fig.~\ref{fig:axis}, where we simultaneously show the identified vortices on three levels $k_n=4$, $16$, and $64$ ($k_n\eta=1.0\times10^{-3}$, $4.0\times10^{-3}$, and $1.7\times10^{-2}$) in the hierarchy. For $h=0$ [Fig.~\ref{fig:axis}(a)], we observe a striking  hierarchy of vortex tubes, whereas for $h=2/3$ [Fig.~\ref{fig:axis}(c)], we only observe blobs of vortices rather than coherent vortex tubes. Since, for small enough $h$, the lower bound of the inequality in (\ref{eq:Qn-cond}) increases with $n$, only strong vortices are extracted. Therefore, it is expected that we observe strong vortices in Fig.~\ref{fig:Viz-Q}(a). However, it is nontrivial that the extracted structures are spatially localized and coherent.

\begin{figure}
\centering
\includegraphics[bb=0 0 650 273,width=0.48\textwidth]{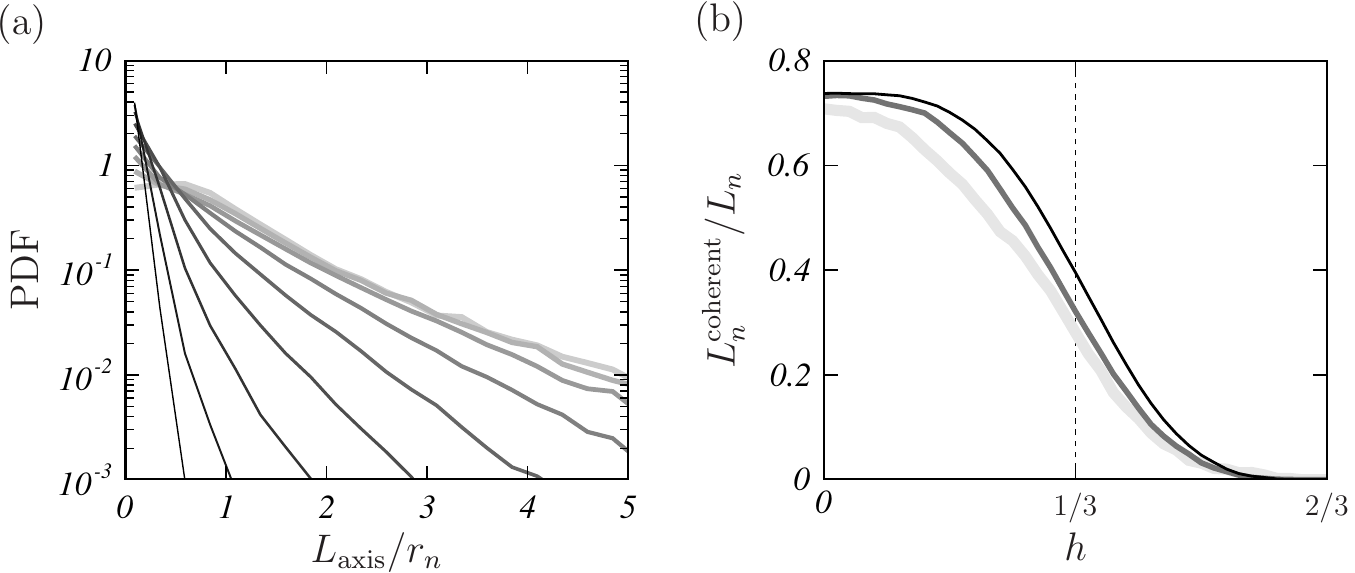}
\caption{(a) PDF of the length $L_\text{axis}$ of each vortex axis in the band of $k_n=64$ for $h=i/60$ ($i=0,5,10,\ldots,40$ from thick gray to thin black lines). (b) The ratio of the length of the coherent vortex axes to the total length for $k_n=16$ (light gray), $32$ (gray), and $64$ (black).}
\label{fig:ratio}
\end{figure}

We are ready to quantify the $h$-dependence of the coherence of vortices observed in Figs.~\ref{fig:Viz-Q} and \ref{fig:axis}. To this end, we show in Fig.~\ref{fig:ratio}(a) the PDF of vortex axis length for $k_n=64$, where different lines correspond to different values of $h$. Only short vortices are observed for larger $h$, whereas longer ones exist for smaller $h$. As shown in the supplementary material, results for other $k_n$ are similar to Fig.~\ref{fig:ratio}(a). Here, we define a coherent vortex as the one longer than $r_n=\pi/k_n$ and estimate the ratio of the sum $L_n^\text{coherent}$ of lengths of coherent vortices to the total length $L_n$ of identified ones. The three lines in Fig~\ref{fig:ratio}(b) show results for $k_n$ in the inertial range. It is clear that the quantitative change of the coherence occurs around $h=1/3$.

Incidentally, estimating the energy flux $\Pi(r)$ on the scale $r$ as $\Pi(r)\sim\Delta u(r)^3/r\sim r^{3h-1}$, we see that structures for $h>1/3$ are irrelevant to the energy cascade in small scales ($r\ll L$), while those for $h\ll1/3$ lead to strong energy transfer. Hence, Fig.~\ref{fig:ratio}(b) gives evidence that coherent structures are responsible for the energy cascade.

It is also important to show in Fig.~\ref{fig:VnDhZp}(c) by the dashed line that the capacity dimensions estimated from thus identified vortex axes for different scales is consistent with $D(h)$ estimated from the power laws of $V_n$ in Fig.~\ref{fig:Viz-Q}(a,b).

The order within turbulence, recognized by experimentalists for long \cite{Kline-1967}, has garnered even greater attention since the advent of numerical simulations \cite{Jimenez-2018}. The relationship between the energy cascade and coherent structures is important but inconclusive \cite{Cardesa-2017}. It is surprising that the structures observed for $h\ll1/3$ [Fig.~\ref{fig:axis}(a)] are similar to the hierarchy of coherent vortices shown in Refs.~\cite{Goto-2008,Goto-2017,Goto-2024} to explain the energy cascade in term of vortex stretching \cite{Tennekes}. However, since other mechanisms of the energy cascade such as strain self-amplification \cite{Johnson-2020} were also suggested, the present method to identify structures for each $h$ will enable us to develop more quantitative arguments on the role of coherent and incoherent structures in the energy cascade.

The development of mathematics \cite{Yoneda-2021} may be also triggered by the concrete multifractal structures presented here. For example, if the solutions to the Navier--Stokes equation exhibit singularities, the coherent structures of the form shown in Fig.~\ref{fig:axis}(a) are likely responsible. Such coherent structures also provide clues for the construction of non-unique solutions to the Euler equation via the convex integration \cite{DeLellis-2013,Isett-2018, Buckmaster-2015}.

{\it Conclusions.} We have verified the multifractal theory \cite{Parisi-1985} of turbulence using numerical data of fully developed turbulence at $R_\lambda=1700$. By applying a condition (\ref{eq:Qn-cond}) to the enstrophy of the band-pass-filtered field, we extract the spatial structures for each H\"older exponent $h$ [Fig.~\ref{fig:Viz-Q}(a--c)], and by evaluating the dimension $D(h)$ from its volume [Fig.~\ref{fig:VnDhZp}(a, b)], we have obtained results consistent with those evaluated indirectly from $\zeta_p$ measurements [Fig.~\ref{fig:VnDhZp}(c)]. The spatial structures for each $h$ are coherent for $h\ll1/3$, and those for small $h$ form the hierarchy of coherent vortex tubes [Figs.~\ref{fig:Viz-Q}(d) and \ref{fig:axis}(a)]. In contrast, structures for $h\gg1/3$ are incoherent and featureless [Figs.~\ref{fig:Viz-Q}(f) and \ref{fig:axis}(c)]. It is interesting that $h=1/3$ seems the boundary between order and disorder in spatial structures [Fig.~\ref{fig:ratio}(b)]. Although quantification remains as a future task, since the energy transfer between scales occurs for $h<1/3$, these findings suggest the relevance of coherent structures for the energy cascade. Furthermore, since $h=1/3$ corresponds to Kolmogorov's mean-field theory \cite{Kolmogorov-1941}, and the relationship with Onsager's \cite{Onsager-1949} conjecture regarding the Euler equation is also intriguing. These observation should lead to new understanding of the physics and mathematics of turbulence.

\vspace*{5pt}
{\it Acknowledgments.} The analyzed data was kindly shared by Prof.~Takashi Ishihara. This study was partly supported by JSPS Grants-in-Aid for Scientific Research (24H00186, 25K01158, JP26H00389). The numerical analyses were conducted using the supercomputer Fugaku through the HPCI System Research Projects (hp230288, hp240278, hp260139) and the Plasma Simulator under the auspices of the NIFS Collaboration Research Programs (NIFS24KISC007, NIFS26KISC034).

\end{document}